# Proposal for a New Test of the Time Independence Of The Fine Structure Constant, α, Using Orthogonally Polarised Whispering Gallery Modes in a Single Sapphire Resonator

Michael Edmund Tobar and John Gideon Hartnett

Frequency Standards and Metrology Research Group, School of Physics, University of Western Australia, Crawley 6009.

A new experiment to test for the time independence of the fine structure constant, α, is proposed. The experiment utilizes orthogonally polarized Transverse Electric and Transverse Magnetic Whispering Gallery Modes in a single sapphire resonator tuned to similar frequencies. When configured as a dual mode sapphire clock, we show that the anisotropy of sapphire makes it is possible to undertake a sensitive measurement from the beat frequency between the two modes. At infrared frequencies this is possible due to the different effect of the lowest phonon frequency on the two orthogonally polarized modes. At microwave frequencies we show that the phonon effect is too small. We show that the Electron Spin Resonance of paramagnetic impurities (such as $Cr^{3+}$) in the lattice effects only one polarization with an $\alpha^6$ dependence. This enables an enhancement of the sensitivity to temporal changes in α at microwave frequencies.

## I. INTRODUCTION

Two major tasks in fundamental physics are the quantization of gravity and the unification of all interactions. All approaches to achieve these tasks lead to deviations from present day physics. For example, violations of the Einstein's Equivalence Principle are possible. Some violations may manifest as spatial and/or time variations of the fundamental physical constants such as the fine structure constant α, and recently attempts to measure these effects have gained considerable attention. One of the most precise tools for testing these theories is the frequency or time standard (clock). The measurement requires the comparison of two clocks with different power law dependence on α [1-3]. Alternative approaches examine astrophysical and geophysical data [4, 5].

A novel method using monolithic Fabry-Perot resonators to test for drift in α was recently proposed[6]. The index of refraction of ionic or molecular crystal depends on α; thus a drift in α will cause a change in the dispersion characteristic of the crystal. In particular [6] analyzed the dependence of optical phonons and electronic transitions of ionic impurities, and remarked that the technique was applicable down to microwave frequencies. In this paper we show that the proposed technique needs to be modified at lower frequencies incorporating infrared to microwave. We show that the first phonon resonance at $10^{13}$ Hz is in inherently low loss at 4 K and may offer the best frequency for this type of measurement. At microwave frequencies we show that the introduction of $Cr^{3+}$ ions to the lattice significantly enhance the dispersion and sensitivity due to an Electron Spin Resonance (ESR) at 11.45 GHz.

Temporal drift in α may be detected by exciting two modes in the same resonator that sample different dispersion. For an isotropic material this may only be achieved at different frequencies. Since sapphire is anisotropic, the dual mode technique [7, 8] may be implemented to realize this experiment. Transverse Electric (TE) and Transverse Magnetic (TM) modes may be excited at nearly the same frequency, and will sample different values of dispersion. This effect may be amplified with the presence of paramagnetic impurities. Paramagnetic impurities result in an anisotropic magnetic susceptibility of the lattice. Thus, if both TE and TM modes are excited simultaneously, a drift in α will be measurable from the beat frequency of the two modes. The proposed experiment is to create a 4 K cooled dual mode frequency standard based on a sapphire resonator. The beat frequency will be stabilized and measured. If a positive measurement is made that suggests a drift in α it will be necessary to verify this by constructing two or more systems to distinguish the result from a systematic drift.

## II. DIELECTRIC RESONATOR DEPEDENCE ON FINE STRUCTURE CONSTANT: MICROWAVE FREQUENCIES TO THE INFRARED

To calculate the dependence of the fine structure constant on the refractive index at optical frequencies, in ref [6] the refractive index was related to the electric susceptibility through the local field model. To model a macroscopic resonator we choose to use the one phonon model, and we relate the permittivity, $\varepsilon_r$, to the electronic susceptibility $\chi$, with the following relation $\varepsilon_r=1+\chi$. This model is well known to describe the relative permittivity accurately in sapphire, in the radio frequency to infrared regions, and at temperatures below 50 K[9-11]. At frequencies above the infrared, the susceptibility is summed over more than one lattice vibrational state, but in this case we only consider the region where the lowest phonon frequency influences the results, which avoids the necessity of a summation of states. In this case the complex permittivity is given by;

$$\mathcal{E}_r(f) = 1 + \frac{\chi_{DC}}{1 - \left(\frac{f}{f_i}\right)^2 - j\frac{f}{f_i Q_i}} \qquad (1)$$

Where $f_i$ is the phonon frequency, $Q_i = \frac{f_i}{\Delta f_i}$ is the dimensionless quality factor of the phonon transition and $\chi_{DC}$ is the DC electronic susceptibility, which was shown to be independent of α [6]. Thus, following the same approach we find that the real part of the permittivity of the material may be expressed as;

$$\varepsilon_r(f,\alpha) = 1 + \frac{\chi_{DC}\left(1 - \left(\frac{f}{\alpha^2 K_i}\right)^2\right)}{\left(1 - \left(\frac{f}{\alpha^2 K_i}\right)^2\right)^2 + \left(\frac{f}{\alpha^2 K_i Q_i}\right)^2} \qquad (2)$$

Here the phonon frequency is of the form, $f_i = \alpha^2 K_i$, where $K_i$ is a constant also shown to be independent of α [6]. Also, we assume that the quality factor of the phonon transition is first order independent of α and frequency independent. The frequency independence will be maintained as long as, $f < f_c$, where $f_c$ is the cut off frequency of the validity of the model, which extends to the infrared[9]. Phonon resonances with $f > f_c$ have an exponential frequency dependence on loss, which still satisfies Kramers Kronig relation, in this case (2) takes a more complex form. This lead Braixmaier et al. [6] to concluded that sampling an ionic crystal near a phonon resonance added too much loss to make use of the high dispersion. On the contrary, in this paper we show that the fundamental phonon resonances in sapphire remains low loss near the high sensitivity region close to the phonon resonance frequency.

By implicitly differentiating (2) it is straightforward to show the following relation;

$$\frac{d\varepsilon_r}{d\alpha} = -2\frac{f}{\alpha}\frac{d\varepsilon_r}{df} \tag{3}$$

Thus we can relate a fractional change in permittivity to a fractional change in fine structure constant by;

$$\frac{\Delta\varepsilon_r}{\varepsilon_r} = -2\mathrm{E}\frac{\Delta\alpha}{\alpha} \tag{4}$$

where E is the normalised dispersion coefficient of permittivity given by;

$$\mathrm{E} = \frac{d\varepsilon_r}{df}\frac{f}{\varepsilon_r} \tag{5}$$

Next we must relate the permittivity change to a frequency change. For a dielectric resonator of resonant frequency, $f_r$, with 100% of its energy confined in the dielectric the simple relationship holds.

$$\frac{\Delta f_r}{f_r} = -\frac{1}{2}\frac{\Delta\varepsilon_r}{\varepsilon_r} \tag{6}$$

Thus by combining (6) and (4) the relationship between the dielectric resonant frequency and the fine structure constant is given by;

$$\frac{\Delta f_r}{f_r} = \mathrm{E}\frac{\Delta\alpha}{\alpha} \tag{7}$$

Thus, it is clear that if two frequencies are measured with different values of E, then the beat frequency will be proportional to changes in the fine structure constant.

### III EXPLOITING THE ANISOTROPIC PERMITTIVITY OF SAPPHIRE

Sapphire is a uniaxial anisotropic material, in general both the real and imaginary parts of the permittivity are anisotropic. The loss tangent, $Tan\delta$, of the material is an important parameter for a dielectric resonator, as the quality factor of the resonance, $Q_r$, is equal to the inverse of $Tan\delta$ if 100% of the energy is stored in the dielectric. The loss tangent is calculated from the ratio of the imaginary and real permittivity, and is given by;

$$Tan\delta(f) = \frac{\dfrac{\chi_{DC}}{Q_i}\dfrac{f}{f_i}}{\left(\dfrac{f}{f_i}\right)^4 + \left(\dfrac{f}{f_i}\right)^2\left(\dfrac{1}{Q_i} - 2 - \chi_{DC}\right) + 1 + \chi_{DC}} \tag{8}$$

This is related to the absorption coefficient by;

$$\mathrm{A}(f) = \frac{2\pi f}{c}\sqrt{\varepsilon_r}\ Tan\delta \tag{9}$$

The loss tangent or absorption of a low loss material is particularly hard to measure. For example at microwave frequencies the Whispering Gallery (WG) mode method was developed and remains the only method capable of measuring loss tangents of order $10^{-6}$ or less[12, 13]. This is particularly true for low temperature measurement near 4 K where the loss tangent is smaller than $10^{-9}$. The WG mode method is such a sensitive method that it enabled the first determination of anisotropy in the loss tangent of uniaxial anisotropic crystals[14]. Near 4 K, paramagnetic impurities also effect the result over a broad frequency range even if the transition frequency is as far away as a few THz, which is the case for titanium impurities[15]. This is because the transitions are thermally excited at cryogenic temperatures. Thus, a measurement of the phonon-induced losses is quite difficult unless a very pure sample has been obtained. We have measured many samples in the microwave region, and the losses due to paramagnetic impurities usually limit the $Q_r$ of the resonance around 4 K to a temperature independent value close to $10^9$. The exception is the most pure grade of HEMEX from Crystal systems. In these samples a $Q_r$ of $10^{10}$ at 12 GHz has been measured, which still exhibit a power law of $T^1$, which means the $Q_r$ limit is not due to paramagnetic impurities [16]. We believed that this limit is due to the dielectric lattice. This result was used in combination with the low temperature permittivity versus frequency measurements made by Lowenstein et. al. [17], to calculate the properties of the phonon resonances at 4K. Fitting this data to (1) and (8), the calculated parameters parallel and perpendicular to the c-axis are given in table 1, and the real permittivity, loss tangent and absorption coefficient are plotted as a function of frequency in figures 1 to 3.

Table I Fit parameters to the dielectric properties of the first phonon resonances in sapphire

| Fit Parameter (~4K) | Parallel to c-axis | Perpendicular to c-axis |
|---|---|---|
| $\varepsilon_{DC} = 1+\chi_{DC}$ | 11.2431 | 9.2129 |
| $f_i$ | $1.3 \times 10^{13}$ | $1.6 \times 10^{13}$ |
| $Q_i$ | $8.1 \times 10^6$ | $6.7 \times 10^6$ |

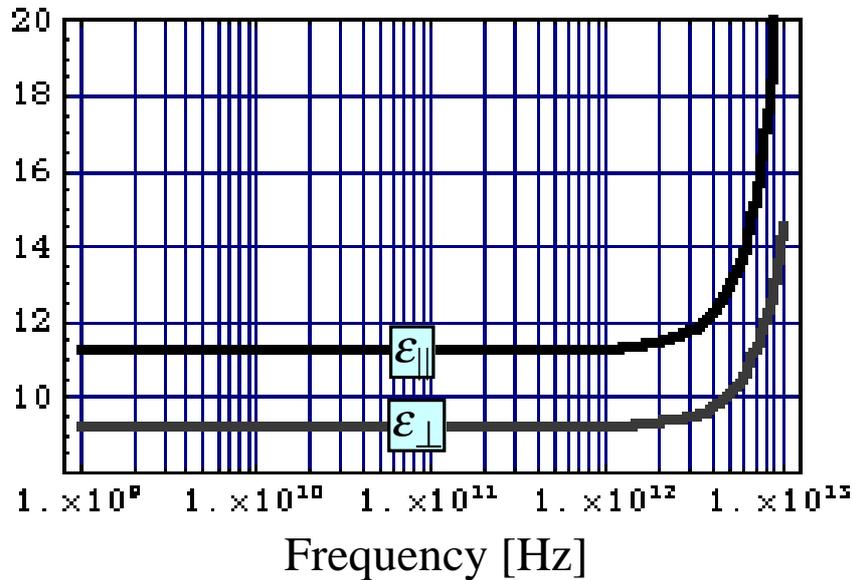

Figure 1.
Real permittivity versus frequency calculated from (1) with the parameters presented in table I. Values are plotted from microwave to infrared for the permittivity parallel, $\varepsilon_\parallel$, and perpendicular, $\varepsilon_\perp$, to the c-axis. Permittivity increases markedly around the first phonon resonance.

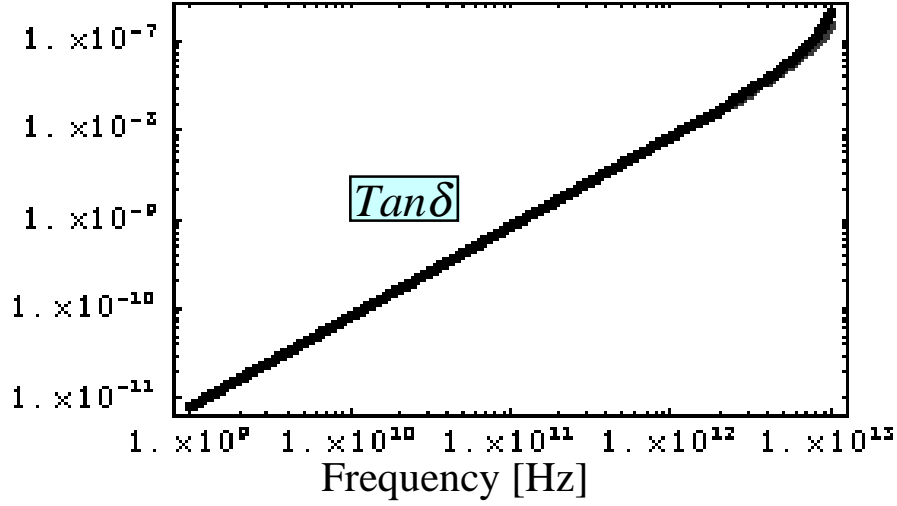

Figure 2
Loss tangent versus frequency calculated from (8) with the parameters presented in table I. Values are plotted from microwave to infrared.

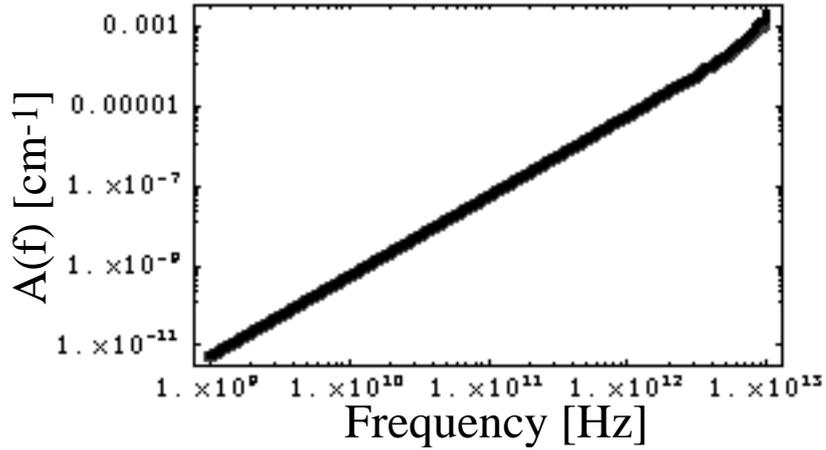

Figure 3
Absorption coefficient [cm$^{-1}$] versus frequency calculated from (9) with the parameters presented in table I. Values are plotted from microwave to infrared.

There is a lack of data on the properties of the first phonon resonance at cryogenic temperatures. However, the DC value of permittivity compares well with excepted values, and the phonon frequency compares well with other published values at room temperature[18]. It is likely that most loss measurements have not been sensitive enough to measure true absorption coefficients or loss tangent at 4 K (except in the case of the WG method), and we intend to verify the calculated losses shown in fig 2 and 3. However we should note that it is well known that the sapphire loss tangent has 1/f dependence as shown.

The idea of this experiment is to exploit the anisotropy of sapphire and construct a Dual Mode interrogation system or oscillator operating on a TE and TM mode at nearly the same frequency, ie $f_{\parallel} \approx f_{\perp} \approx f_r$[8]. The beat frequency can then be monitored to test for the fine structure constant. For this case the sensitivity of the beat frequency to a drift in fine structure constant is calculated from (7) to be;

$$\frac{|\Delta f_{\parallel} - \Delta f_{\perp}|}{f_r} = \mathrm{E}_{diff} \frac{\Delta \alpha}{\alpha}, \text{ where } \mathrm{E}_{diff} = |\mathrm{E}_{\parallel}(f) - \mathrm{E}_{\perp}(f)| \qquad (10)$$

Figure 4 shows the sensitivity function, $E_{diff}$, as a function of frequency.

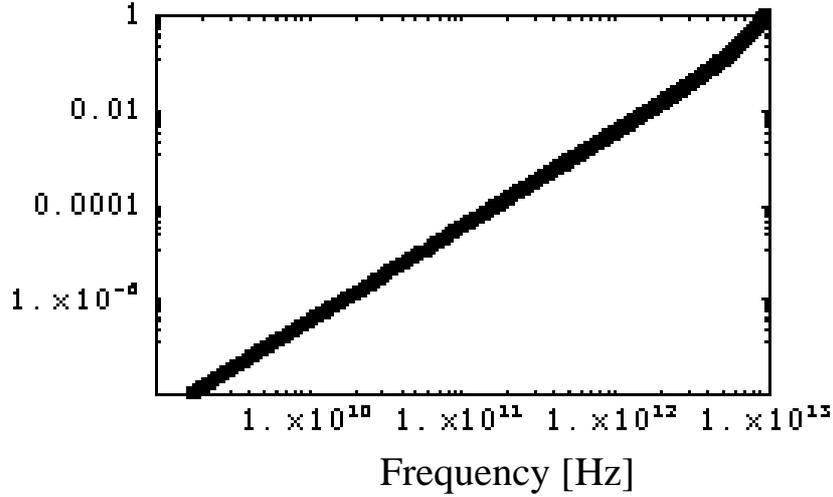

Frequency [Hz]

Figure 4

Calculated $E_{diff}$, as a function of frequency calculated from (1) and (5) with the parameters presented in table I

Because the modes are excited in the same crystal, some form of common mode rejection of noise sources could be expected as discussed in [6]. Also, the difference frequency will be much lower than the resonance frequency. This down conversion will lower the noise requirements of the readout of the beat frequency. The key point to this analysis is that close to the phonon frequency a sensitivity function of order unity can be obtained with low loss. One would like to excite modes in the sensitive frequency range of $10^{12}$ to $10^{13}$ Hz (300 to 30 μm wavelengths). The frequency of a pure WG mode to first approximation is given by;

$$f_r = \frac{c\,m}{2\pi r \sqrt{\varepsilon}} \quad (11)$$

Where c is the speed of light, m is the azimuthal mode number, r is the sapphire radius and ε is either the permittivity parallel (for a TM mode) or permittivity perpendicular (for a TE mode). For example, orthogonal WG modes could be excited in a 5 mm diameter sapphire cylinder with m varying from 160 to 2600, and a Free Spectral Range of 4 to 6 GHz. Single mode sapphire clocks with low fractional frequency instabilities of order $10^{-16}$ to $10^{-15}$ have been already demonstrated at microwave and optical frequencies[19, 20]. This method should be possible to adapt to the suggested frequencies and has the potential for high sensitivity tests of drifts in the fine structure constant of order $10^{-15}$/yr as discussed in [6].

## IV. PARAMAGNETICALLY DOPED SAPPHIRE

At microwave frequencies paramagnetic impurities supply transitions between electron spin states. This changes the susceptibility of the sapphire lattice, which is typically an anisotropic effect. The magnitude of the susceptibility and the frequency of the Electron Spin Resonance (ESR) depend on the fine structure constant. In the following we derive this dependence and calculate the influence on the frequency of a dielectric resonance in the sapphire crystal due to the influence of the $Cr^{3+}$ ESR transition in sapphire.

### A. Determination of the Dispersion Added by $Cr^{3+}$ ions

Paramagnetic $Cr^{3+}$ ions in sapphire only change the magnetic susceptibility perpendicular to the c-axis. The form of the complex susceptibility added by the ESR transition may be written as;

$$\chi_\perp(f) = \chi'(f) + j\chi''(f) = \frac{\chi_\perp^{DC}}{\left(1-\left(\frac{f}{f_0}\right)^2\right)^2 + \left(\frac{f}{f_0}\frac{1}{Q_0}\right)^2}\left[\left(1-\left(\frac{f}{f_0}\right)^2\right) + j\left(\frac{f}{f_0}\frac{1}{Q_0}\right)\right] \quad (12)$$

Here $\chi_\perp^{DC}$ is the DC magnetic susceptibility, $f_0$ is the frequency of the ESR (11.45 GHz for $Cr^{3+}$ in sapphire[21]), $Q_0 = \frac{f_0}{\Delta f_0} = \pi f_0 \tau_0$, where $\Delta f_0$ is the bandwidth of the ESR and $\tau_0$ is the relaxation time of the ESR. The loss added to the lattice can be characterized by the magnetic loss tangent, $\tan\delta_m$ and is given by;

$$\tan\delta_m(f) = \frac{\chi''}{1+\chi'} \approx \frac{\frac{\chi_\perp^{DC}}{Q_0}\frac{f}{f_0}}{\left(1-\left(\frac{f}{f_0}\right)^2\right)\left(1+\chi_\perp^{DC}-\left(\frac{f}{f_0}\right)^2\right)} \quad (13)$$

To evaluate the magnetic susceptibility we use data taken at the University of Western Australia by Mann and analyzed by Krupka[22]. The experiment consisted of analyzing the WG modes in a 5cm diameter sapphire cylinder with a few parts per million impurity ions of $Cr^{3+}$, enclosed in a 8cm diameter copper cavity. The measured real part of the susceptibility is shown in figure 5. Mann and Krupka fitted the data with $\chi_\perp^{DC}$ and $\tau_0$ as free parameters and calculated $\chi_\perp^{DC}=5.1\times10^{-8}\pm2\%$ and $\tau_0=0.69\times10^{-8}\pm6\%$ seconds. We point out here that fitting to the real part of the permittivity can not give an accurate determination of the ESR relaxation rate or bandwidth as a fit to the real part is very insensitive to this parameter. Thus we believe the quoted errors are likely to be wrong, especially so for $\tau_0$. To obtain a more accurate fit the magnetic loss tangent must be calculated. To calculate this effect, WG mode Q-factors, $Q_r$, must be measured and the magnetic filling factor perpendicular to the c-axis, $p_{m\perp}$, calculated. In general the $Q_r$ of a TM dielectric resonance is given by;

$$Q_r^{-1} = p_{m\perp}\tan\delta_m + \frac{R_s}{G} \quad (14)$$

Here $R_s$ is the surface resistance of the Copper cavity and G is the geometric factor of the mode. Thus to calculate $\tan\delta_m$ one needs to measure Transverse Magnetic modes with $p_{m\perp}$ close to unity and a high G-factor. The highest G-factor TM modes are the fundamental $WGH_{m,0,0}$ mode family. Only these modes are selected from the measured modes of Mann and Krupka[22] shown in figure 5, as all other modes are either limited by the copper cavity or have a low sensitivity to the perpendicular magnetic field. The Q-factors and filling factors of the measured $WGH_{m,0,0}$ modes are shown in table II.

Table II $WGH_{m,0,0}$ mode family data

| Mode | Frequency | $p_{m\perp}$ | G-factor | Q-factor |
|---|---|---|---|---|
| $WGH_{12,0,0}$ | 8.95045 | 0.881 | $1.63\times10^7$ | $1.8\times10^8$ |
| $WGH_{13,0,0}$ | 9.56922 | 0.888 | $3.92\times10^7$ | $1.0\times10^9$ |
| $WGH_{14,0,0}$ | 10.1862 | 0.893 | $9.42\times10^7$ | $1.3\times10^9$ |
| $WGH_{15,0,0}$ | 10.80142 | 0.898 | $2.26\times10^8$ | $2.2\times10^8$ |

If the data was only limited by the magnetic loss tangent one would expect an increasing Q-factor the further from the 11.45 GHz resonance. This is not the case due to the cavity losses effecting the measurement due to smaller confinement for lower m numbers. Figure 6 clearly shows this effect, ignoring the $R_s$, the effective $\tan\delta_m$ is plotted as a function of frequency. The two points closest to the 11.45 GHz resonance are identified to be limited by the magnetic loss tangent, while the two low frequency points are limited by the copper surface resistance. Hence by undergoing a simultaneous fit to the real susceptibility and the loss tangent, a more accurate determination of the ESR parameters can be obtained. We calculate $\chi_\perp^{DC}=5.9\times10^{-8}\pm6\%$ and $\Delta f_0 = 9\times10^6$, which gives a $\tau_0=3.5\times10^{-8}$ seconds. Note that the DC susceptibility is very close to Mann and Krupka's calculation, however the spin-spin relaxation rate was calculated to be a factor of 5 longer.

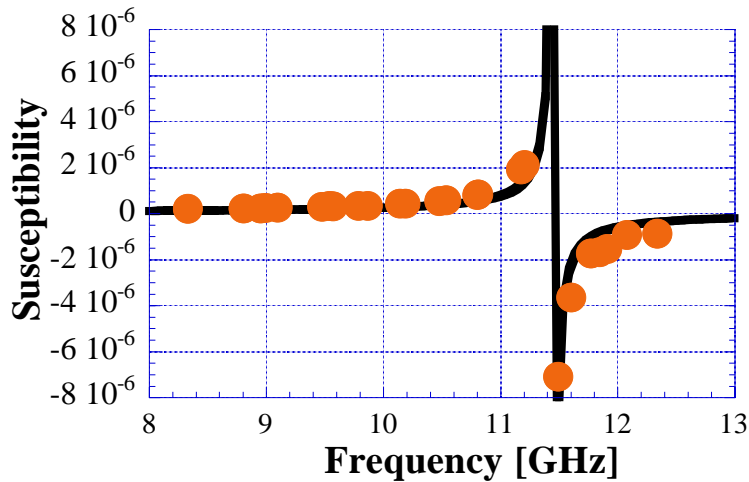

Figure 5

Magnetic DC susceptibility versus frequency. Dots show calculations due to frequency shifts in WG modes (data taken from Mann and Kruka[22]), while the bold line shows the best fit from (12) and (13).

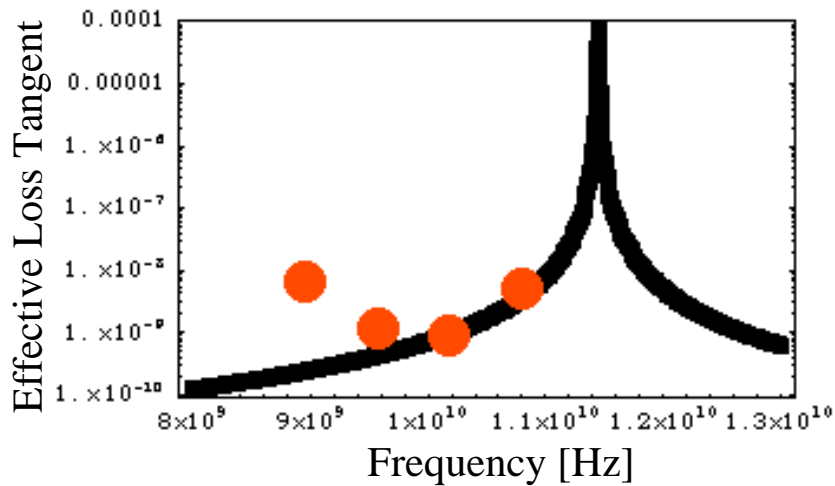

Figure 6

The magnetic loss tangent versus frequency in Hz. Dots show calculated effective loss tangent due to Q-factors of the $WGH_{m,0,0}$ mode family, which is calculated assuming Rs =0 in (14). The bold line shows the best fit from (12) and (13). Note the two lower frequency calculations diverge form the predicted due to the smaller G-factors, which make the measurements limited by the surface resistance of the copper cavity. We have calculated from (14) that a surface resistance of the order of 50 mΩ explains this effect.

In the case of magnetic spin-spin transitions, we show later that unlike the electronic case calculated previously[6], the DC susceptibility depends on $\alpha$. Thus we take a more general approach to calculating the sensitivity. First we must relate the real magnetic susceptibility change to a frequency change. For a dielectric resonator of resonant frequency, $f_r$, with 100% of its energy confined in the dielectric, the simple relationship holds.

$$\frac{df_r}{d\chi'} = -\frac{1}{2}\frac{f_r}{1+\chi'} \tag{15}$$

Using implicit differentiation we can show the following relationship holds;

$$\frac{\Delta f_r}{f_r} = M\frac{\Delta \alpha}{\alpha} \quad \text{where } M = \frac{\alpha}{f_r}\frac{df_r}{d\chi'}\frac{d\chi'}{d\alpha} \tag{16}$$

Here M is the sensitivity conversion of the mode frequency with respect to fine structure constant. By combining (15) and (16) it follows that;

$$M = -\frac{1}{2}\frac{d\chi'}{d\alpha}\frac{\alpha}{1+\chi'} \tag{17}$$

To calculate the sensitivity conversion (17) must be evaluated over all parameters that vary with $\alpha$. In the following analysis we show that the DC susceptibility and ESR frequency are highly dependent on $\alpha$, and we do not consider any dependence on the spin-spin relaxation time as this will have no first order effect on the frequency of the ESR transition or the dielectric resonance.

## B. $Cr^{3+}$ ESR Frequency Dependence on $\alpha$

The $Cr^{3+}$ ion has three unpaired electrons, with the free-ion ground state of $^4F_{3/2}$. When the ion is present in a tetragonal lattice like sapphire, the non-cubic symmetry of the field causes a zero applied field splitting of the spin states (or Electron Spin Resonance) due to the crystal field. Orton has used crystal field theory [23] to calculate the energy associated with the transition between spin states using second order perturbation theory. The zero field splitting was calculated to be;

$$\Delta E = E_{\pm 3/2} - E_{\pm 1/2} \approx \frac{8\delta\lambda^2}{\Delta^2} \tag{18}$$

Here $\lambda$ is the spin-orbit coupling, $\Delta$ is the effective isotropic field splitting between the ground ($^4A_2$) and first excited state ($^4T_2$) of the $^4F$ orbital states and $\delta$ is the splitting of the first excited state due to the anisotropic crystal field. To calculate crystal field effects, the potential experienced by an electron wave function due to the nearest neighbor electrons is calculated[23]. From this calculation one can show the following.

$$\Delta = X_\Delta \frac{e^2}{4\pi\varepsilon_o}\frac{\langle r^4 \rangle}{a^5} \quad \text{and} \quad \delta = X_\delta \frac{e^2}{4\pi\varepsilon_o}\frac{\left(1-\frac{b}{a}\right)\langle r^4 \rangle}{a^5} \tag{19}$$

Here $a$ is the atomic spacing perpendicular to the c-axis, $b$ is the atomic spacing parallel to the c-axis, $\langle r^4 \rangle$ is the mean radius to the power of four of the electron wave function and $X_\Delta$ and $X_\delta$ are dimensionless constants of order unity. Now we use similar reasoning to Braixmaier et. al. [6], that the inter-atomic spacing and the wave function size is proportional to the Bohr radius, $a_o$, of the electron, and by substituting (19) into (18) we obtain;

$$\Delta E \propto \lambda^2 \frac{a_o 4\pi\varepsilon_o}{e^2} \tag{20}$$

The spin-orbit coupling and Bohr radius may be written in terms of the fine structure constant by;

$$\lambda = m_e c^2 Z^4 \alpha^4 \quad \text{and} \quad a_o = \frac{\hbar}{\alpha m_e c} \tag{21}$$

By substituting (21) into (20) one may show that the frequency of the transition is given by:

$$f_o \propto \left(\frac{c}{\Lambda}\right) Z^8 \alpha^6 \quad \Lambda = \frac{h}{m_e c} \tag{22}$$

Here $\Lambda$ is the Compton wavelength. Thus the frequency of the paramagnetic resonance is proportional to the sixth power on $\alpha$. Experimentally the resonance has been confirmed to be at 11.45 GHz, and thus a microwave sapphire clock operating close to this frequency is suited for this type of experiment.

## C. $Cr^{3+}$ ESR Susceptibility Dependence on $\alpha$

The susceptibility of the transition is proportional to the DC susceptibility given by [21];

$$\chi_\perp^{DC} \propto N \mu_B^2 \quad N \sim \frac{1}{a_o^3} \quad \mu_B \sim \frac{e\hbar}{2m_e c} \tag{23}$$

Here N is the number density of ions, k is Boltzmann's constant and $\mu_B$ is the Bohr magneton. Thus, it is straightforward to show that;

$$\chi_\perp^{DC} \propto \alpha^4 \tag{24}$$

Unlike the case of a phonon resonance, the DC susceptibility due to the spin-spin interaction is dependent on the fine structure constant.

## D. Sensitivity of Cr3+ Doped Sapphire to $\alpha$

Braxmaier et. al. [6] considered electronic transitions in doped ionic crystals to increase the electronic dispersion in the crystal. They assumed that the normalized sensitivity (in our case M) was unity, in the following we do some similar analysis for doped sapphire at microwave frequencies. By combining (22) and (24) with (17) the normalized sensitivity of a dielectric mode in doped sapphire is given by;

$$\mathrm{M}(f) = \frac{3f \frac{\partial \chi'}{\partial f} - 2\chi'}{1 + \chi'} \tag{25}$$

For the doped sapphire investigated in this paper, the sensitivity and inverse magnetic loss tangent as a function of frequency is plotted in fig 7 and 8. The sensitivity is unity at the resonance frequency with a loss tangent of order $10^{-5}$. The sensitivity could be further enhance with the deliberate addition of more Cr3+ ions (ruby). However, the enhancement in sensitivity compromises the loss tangent of the material. This means a better lock to the resonance is needed for the experiment to work. Another option is to find a transition in sapphire with a longer relaxation time.

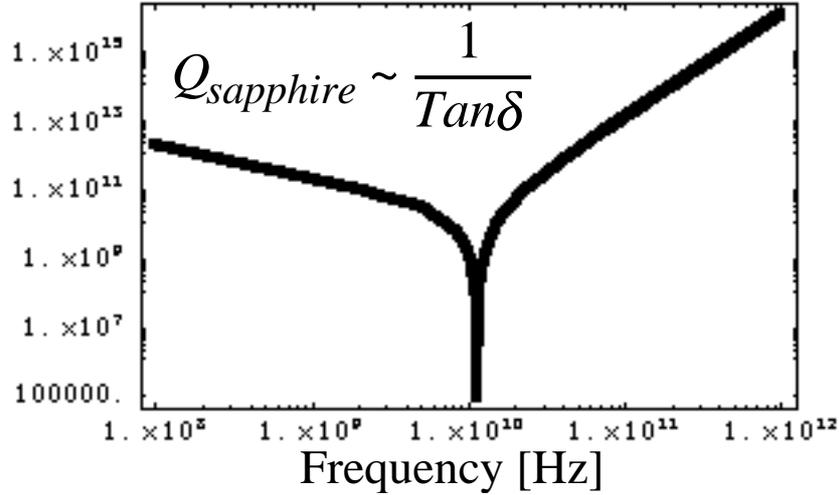
Figure 7
Calculated magnetic Q-factor limit versus frequency for a TM mode in the doped sapphire presented in this paper.

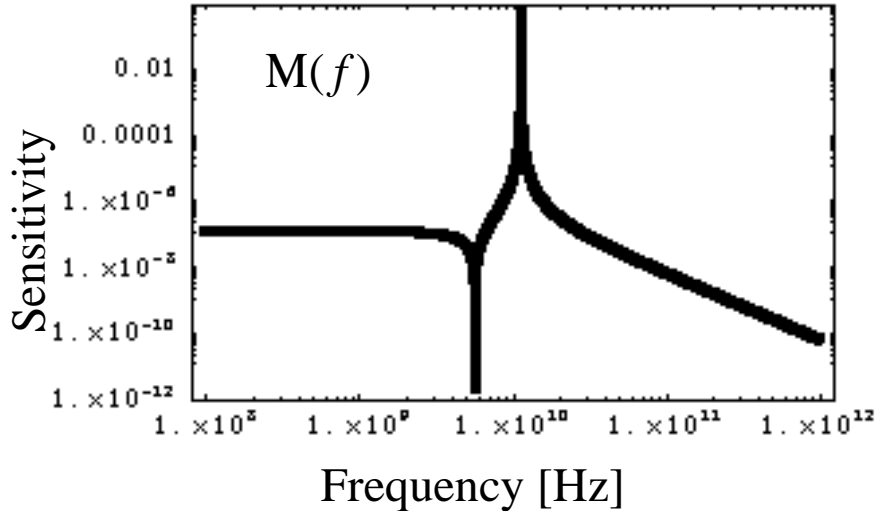
Figure 8
Calculated sensitivity, M, as a function of frequency using (25) and (12) for the doped sapphire presented in this paper

**V. DISCUSSION**

The simple requirement for this technique requires a low-loss crystal resonator with large dispersion. The two requirements are in fact contradictory due to the Kramers-Kronig relations between the real and imaginary part of a resonance system. We have shown that the first phonon resonance at $10^{13}$ Hz is in inherently low loss at 4 K and may offer the best frequency for this type of measurement. At microwave frequencies zero field splitting of the ESR due to $Cr^{3+}$ ions in the sapphire lattice significantly enhances the dispersion and hence sensitivity. However we show that due to Kramers Kronig relation $Cr^{3+}$ adds magnetic losses due to a short spin-spin relaxation time of $3.5\times10^{-8}$ s. Improvement of this technique requires the addition of paramagnetic impurities with longer spin-spin relaxation times.

Braxmaier et. al. [6] showed a similar technique at optical frequencies is capable of measuring drifts in the Fine Structure Constant of the order $10^{-15}$/year. We have proposed two new experiments that can achieve a similar result at lower frequencies.

## ACKNOWLEDGMENTS

The authors would like to thank Dr. Anthony Mann for supplying the Q-factor data in supplement to the already published data by Mann and Krupka. This work was funded by the Australian Research Council.